# Toward a Real-Time Digital Twin Framework for Infection Mitigation During Air Travel

Ashok Srinivasan
University of West Florida
11000 University Parkway
Pensacola, FL 32314
asrinivasan@uwf.edu

Satkkeerthi Sriram
University of West Florida
11000 University Parkway
Pensacola, FL 32314
ss376@students.uwf.edu

Sirish Namilae
Embry-Riddle Aeronautical University
1 Aerospace Blvd
Daytona Beach, FL 32114
namilaes@erau.edu

Andrew Mahyari
UWF/IHMC
15 SE Osceola Ave
Ocala, FL 34471
amahyari@uwf.edu

*Abstract*— Pedestrian dynamics simulates the fine-scaled trajectories of individuals in a crowd. It has been used to suggest public health interventions to reduce infection risk in important components of air travel, such as during boarding and in airport security lines. Due to inherent variability in human behavior, it is difficult to generalize simulation results to new geographic, cultural, or temporal contexts. A digital twin, relying on real-time data, such as video feeds, can resolve this limitation. This paper addresses the following critical gaps in knowledge required for a digital twin. (1) Pedestrian dynamics models currently lack accurate representations of collision avoidance behavior when two moving pedestrians try to avoid collisions. (2) It is not known whether data assimilation techniques designed for physical systems are effective for pedestrian dynamics. We address the first limitation by training a model with data from offline video feeds of collision avoidance to simulate these trajectories realistically, using symbolic regression to identify unknown functional forms. We address the second limitation by showing that pedestrian dynamics with data assimilation can predict pedestrian trajectories with sufficient accuracy. These results promise to enable the development of a digital twin for pedestrian movement in airports that can help with real-time crowd management to reduce health risks.

## TABLE OF CONTENTS

## 1. INTRODUCTION

Pedestrian dynamics studies the movement of individuals within crowds or in pedestrian-dense environments [1]. Pedestrian dynamics models help simulate the trajectories of pedestrians to examine different "what-if" scenarios, which can be used for decision making, such as in the design of the built environment or policies for safe and efficient movement of people. For example, it has been used to evaluate the future capacity of the New York Staten Island Ferry [2] and reduce the risk of stampedes in crowded situations, such as concerts, building evacuations, stadiums, or commercial centers [3].

This paper is motivated by its application to air travel. In past work, it has been used to study infection spread on airplanes and its mitigation by employing suitable boarding procedures [4-7] and in the design of airport security queues [8, 9]. In these studies, trajectories of pedestrians under different boarding policies and queue designs respectively were used to estimate how close passengers get to each other. This proximity of passengers was then linked to an infection model to estimate infection spread of contact-based diseases.

Various mathematical and computational methodologies have been developed to simulate pedestrian movements, each with a unique approach to capturing the complexities of human behavior. Cellular Automata (CA) represent pedestrian movement on a grid, where each cell corresponds to a possible position [2, 10]. The Lattice Gas model also represents pedestrians on a discrete grid, with each cell being either occupied or empty. However, it differs by emphasizing fluid-like, probabilistic movement and collision avoidance, which allows for the simulation of emergent flow patterns and collective behaviors within pedestrian crowds [11].

Agent-Based models treat each pedestrian as an autonomous agent, with each agent following a set of rules in responding to changes in the environment [12]. Social force models describe the motion of individuals through social forces [1]. Typically, propulsive forces motivate a pedestrian towards its goals, while repulsive forces model impediments to this, such as a repulsive force that models slowing down due to others in front of the pedestrian. This is mathematically represented by an ordinary differential equation (ODE), which is numerically integrated to obtain pedestrian trajectories.

Social force models have been used to identify interventions to mitigate infection risk in air travel, such as design of boarding and disembarkation procedures and the design of airport security queues. Aspects of human behavior that are not directly captured in the social force model are either explicitly coded or inferred from data sources, such as location based services data [8].

A significant challenge to identifying effective infection mitigation in air travel through pedestrian dynamics lies in the inherent unpredictability of human behavior, which varies with context and culture. Consequently, generalizing the models, developed in one context, to another context is difficult. In addition, the response of pedestrians to new interventions is difficult to predict.

For effective real-time crowd management, it is crucial to accurately predict and respond to pedestrian behavior. A digital twin (DT) for pedestrian dynamics is a promising solution for this problem. A digital twin will update the pedestrian dynamics model continuously based on real-time data, such as video feeds, from the airport. This will inform a decision layer that suggests suitable interventions, which subsequently impact the pedestrian movement. This allows for a feedback loop of predictions of pedestrian behavior and interventions to manage pedestrian behavior effectively to mitigate infection risk. Our vision for such a digital twin is further explained in Section 2.

There are certain gaps in knowledge to develop such a digital twin. This paper's goal is to generate the following essential knowledge that can subsequently lead to the generation of an effective digital twin for the use of pedestrian dynamics in air travel.

One limitation of the current state of the art in social force models lies in inadequately capturing collision avoidance when two pedestrians are moving on a path toward collision. Collision avoidance behavior can generally be classified into two main types: stopping and swerving [13]. Swerving, which is the change of speed and direction of the pedestrian to go around the other pedestrian, is not modeled accurately. We address this in Section 3.

Another limitation is the lack of literature on updating social force models to account for the departure of observed pedestrian behavior from the one predicted by the model. While data assimilation is well studied for physical processes, it is not known whether techniques designed for physical systems are effective for pedestrian behavior. We address this in Section 4.

Solutions to the above two limitations will enable a future digital twin for pedestrian dynamics to accurately model human movement, using real-time updates video feeds. These can be input into existing infection spread models to identify effective real-time interventions to mitigate infection spread in airports, such as announcements to modify pedestrian behavior or the installation of soft barriers to guide their movement.

We use video data on pedestrian movement to develop suitable models and a different set of videos to evaluate them. Our results show the effectiveness of our solutions to both problems, thus promising to enable the development of a pedestrian movement digital twin for airports that can help with real-time crowd management to reduce health risks.

## 2. DIGITAL TWIN VISION

In this section, we explain our vision for a pedestrian movement digital twin for infection mitigation in airports. Its architecture is provided in Figure 1 below.

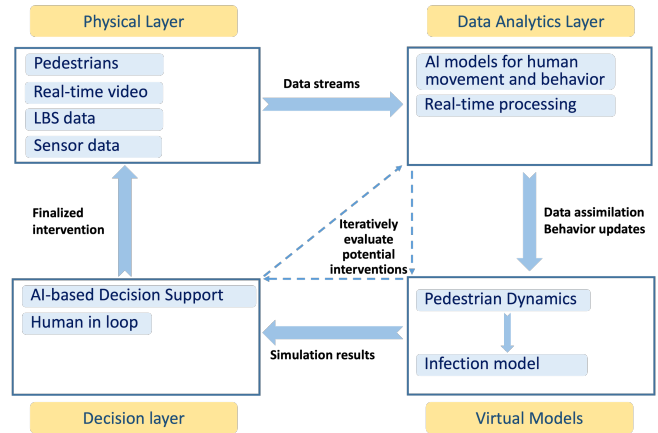

**Figure 1. Architecture of a future digital twin**

We consider an airport where decision makers wish to deploy a digital twin to help mitigate infection risk. The physical system would primarily consist of passengers moving in the airport. Airport management can, additionally, deploy CCTV to observe the movement and behavior of pedestrians and other types of sensors, such as thermal scanning infrared cameras to note any increase of temperature, which may indicate persons with fever. In addition, the airport decision makers could purchase location based services (LBS) data from vendors to provide additional information to the DT. Within the scope of this paper, we consider only human movement information collected from video data.

The above data streams would be sent to a real-time processing module of the data analytics layer. This module would extract features useful to the DT, such as passenger movements patterns and relevant behavior, such as mask wearing. It would also perform data assimilation to update the pedestrian dynamics model. New behavioral features observed by this module can also be added to the pedestrian dynamics model. We have developed the VIPRA pedestrian dynamics cyberinfrastructure that enables such behavior to be included at run time.

The data analytics layer will also include modules with AI-based models for human behavior and movement. This would have been trained on historical data under different circumstances to predict human response to different types of interventions. By relating the observed human behavior and movement to the models trained on historical data, it will try to predict the response to interventions. Note that the impact of some interventions, such as the impact of masks, can be computed theoretically based on the mask filtration efficiency, while some, such as the probability that someone would wear a mask, would be estimated by the AI model.

Others, such as the impact of announcements, would involve a prediction of human behavior by the AI model and its subsequent use by the pedestrian dynamics model to determine movement patterns, and finally their use by the infection model to estimate infection spread likelihood in the virtual layer.

The results from the virtual layer would be sent to the decision layer, which we envision as including a human in the loop. The AI-based decision support system in this layer would propose different interventions, steered by the human), with different possible infection outbreak scenarios iteratively produced by the virtual layer based on input from the data analytics layer. The decision layer will produce a final decision taken by the human, which would then impact the passengers in the physical layer, completing a feedback loop that would repeat.

The goal of this paper is to generate knowledge for realistic simulation of pedestrian movement currently lacking in social force models (on collision avoidance) and on identifying the effectiveness of data assimilation techniques developed for physical systems to human movement.

## 3. COLLISION AVOIDANCE

We first address the accurate modeling of collision avoidance when two pedestrians are moving on a path toward collision. At least one of the pedestrians may stop or swerve to avoid collision. The latter includes the possibility of changing trajectory and speed. – changes in speed and trajectory

Stopping behavior is modeled satisfactorily by social force models. Current state of the art social force models address swerving only when one person is stationary. The SPED [4, 5] and CALM [6, 7] models, which have been used for simulations related to air travel, were used in contexts that don't require swerving, such as airplane boarding and airport security lines. There is a significant gap in current models due to their inability to adequately simulate collision avoidance trajectories when both pedestrians are in motion [10, 14, 15]. Addressing this limitation is crucial for achieving realistic simulations in airports, where such contexts frequently arise.

Such collision avoidance has been studied with other modeling approaches. For example, Liang et al. [10] used a cellular automaton to study collision avoidance in classroom evacuation. Zeng et al. [16] introduced a virtual force field-based (QVFF) navigation algorithm to ensure that robots don't block humans' paths.

There have also been empirical studies on collision avoidance. Parisi et al. [13] performed controlled experiments to analyze pedestrian avoidance behavior with 20 volunteers in an empty parking lot, tracked via a GoPro camera. Their trajectories were recorded within a 5×2.8 m² area at 30 frames per second. Multiple scenarios were considered, such as one pedestrian avoiding a stationary pedestrian, head-on and perpendicular avoidance between two pedestrians, and perpendicular and counterflow configurations of small and large groups. We use data from these experiments in our modeling.

Huber et al. [14] performed experiments with a non-reactive interferer to observe the collision avoidance behavior of a pedestrian. The study investigates how pedestrians avoid collisions with other pedestrians by adjusting their walking path, speed, or both, and how these adjustments depend on the angle and speed of approach. Path adjustments were observed in all scenarios while speed adjustments were observed only in cases where the angle of approach was acutely angled. In this paper, we focus on intersections at roughly perpendicular directions. Thus, we expect no significant change in speed, which was supported by the data that we used.

Olivier et al. [15] explore how pedestrians avoid collisions when walking along crossing paths and how each pedestrian contributes to avoiding collisions. Thirty participants (11 women and 19 men, average age 26.1 years) took part in the study within a 15x15 meter experimental area. Adaptations in speed and orientation to avoid collisions were quantified by their effect on the Minimal Predicted Distance (MPD), which is the predicted closest future distance between two pedestrians on their current paths. It found out that the pedestrian giving way makes most of the adjustment. We use insights from this work. In particular, we found that the use of MPD as a variable leads to simple and accurate models for swerving.

We used the Pedestrian Dynamics Data Archive of the Juelich Research Centre from the study reported in [13]. We considered only scenarios where only two pedestrians cross each other roughly perpendicularly, in order to model the most frequent case. We scaled the pixel data to yield data in centimeters, and rotated the paths so that they were parallel to the x and y axes respectively. These steps help us to derive an expression that is independent of the video resolution or absolute positions of the pedestrians.

We also smoothed the data by using a moving average smoothing algorithm, averaging data points within a window size of 30 time steps.

We chose the region to model as follows. We determine the point of minimum approach distance, which is the shortest distance between two pedestrians throughout their entire trajectory. The region of approach encompasses all data points collected one second before and one second after this point. The direction of motion of the swerving pedestrian is taken as the x axis. Thus, any deviation from the direction of motion is indicated by the value of y.

The model for swerving, consequently, expresses y as a function of other variables. We found that an expression of the following form fits all the different curves.

$$y = a_0 + a_1 x + a_2 D + a_3 x^2 + a_4 D^2 + a_5 xD, \quad (1)$$

where D is the current Euclidean distance between the two pedestrians. D is not independent of x and y. In a predictive model, we use the D value at the previous time step to approximate D. Including D results in a simple multivariate quadratic fit that works for all observed curve shapes. Figures 2 and 3 illustrate this. Note that the parameter values vary for each encounter. Given the common functional form, we can choose parameters based on the distribution of observed parameter values if we wish to simulate typical swerving behaviors.

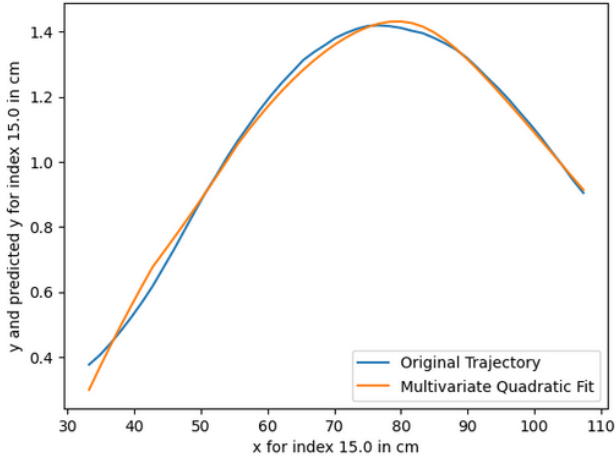

**Figure 2. Multivariate Quadratic fit for trajectory in the region of approach for pedestrian #15. Here, $a_0$ = -5.21, $a_1$ = 0.172, $a_2$ = -0.036, $a_3$ = -0.0011, $a_4$ = 0.009, $a_5$ = 0.0002.**

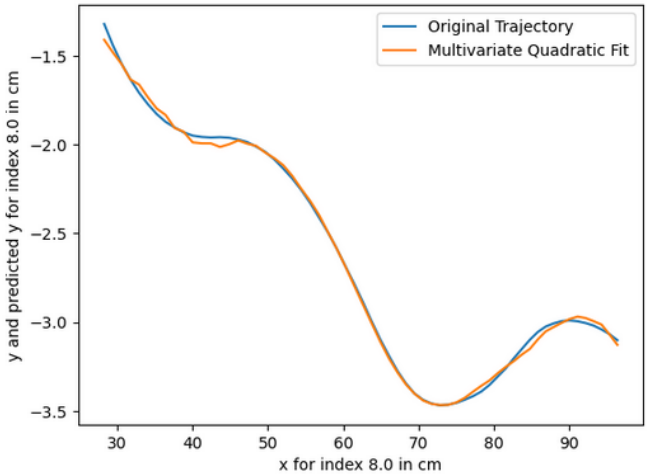

**Figure 3. Multivariate Quadratic fit for trajectory in the region of approach for pedestrian #8. Here, a0 = 319.9, a1 = -9.82, a2 = 0.031, a3 = 0.075, a4 = -0.076, a5 = 0.0003.**

For digital twin purposes, though, we would like to predict the trajectories of passengers currently being observed. This is addressed in Section 4.

## 4. DATA ASSIMILATION

We now address dynamic predictions of trajectories. We start out with a common data assimilation technique – Kalman filter. However, it has certain limitations for our application. (i) In order to deal with uncertainty in human behavior, we typically generate several trajectories and evaluate interventions against these possible scenarios. Kalman filter, on the other hand, yields one predicted trajectory. (ii) Decision making needs a human in the loop, and it will be useful for the human to understand any fundamental behaviors that are changing, rather than just obtaining predictions.

We approach the above challenges in two alternate ways. One is to perform data assimilation to correct the parameters of Equation 1 as described below. The other is to use symbolic regression to identify new functional forms and parameters for the swerving behavior, as an alternative to the form shown in Equation 1.

Symbolic regression is a type of machine learning that finds easy-to-understand equations from data. The aim is to produce clear equations that show how dependent variables relate to the independent variables in the data. This can be useful in new contexts, where pedestrian behavior differs from that used in the training data. There are several tools for performing symbolic regression, such as Operon, DSR, EQL, QLattice, SR-Transformer, and PySR [17]. We evaluated several tools on sample datasets and decided to use PySR.

PySR utilizes a distributed backend implemented in Julia, called SymbolicRegression.jl, complemented by a search algorithm [17]. At its core, PySR employs a multi-population evolutionary approach, iterating through an evolve-simplify-optimize cycle. This process includes evolving populations of expressions through mutations and crossovers, periodically simplifying these expressions using algebraic equivalencies, and refining numerical constants in the expressions using gradient-based optimization techniques. Additionally, PySR incorporates an adaptive parsimony metric that adjusts the complexity penalty dynamically throughout the search.

We first compare the predictions from an Unscented Kalman filter against the following simple dynamic update strategy for Equation 1:

$$y_p(t) = y_e(t) + y(t_i) - y_e(t_i), \quad (2)$$

where $y_p$ is the final prediction trajectory, $y_e$ is the initial estimate for the trajectory based on extrapolation using Equation 1, $t_i$ is the last timestep where the actual trajectory has been observed, and *y* is the observed trajectory. The corrections in Equation 2 are performed every 5 time steps.

We also need an initial prediction, before there is data to perform the correction in Equation 2. We use the following strategy. Data for the pedestrian position, velocity, and distance to the colliding pedestrian from the training dataset is used to estimate the parameters for Equation 1, and the partial trajectories for the first 10 times steps are calculated from this.

We can see from Figures 4 and 5 that the initial predictions are not very accurate, but the update scheme quickly corrects for it. Even in the initial phase, the error is only around 5 cm.

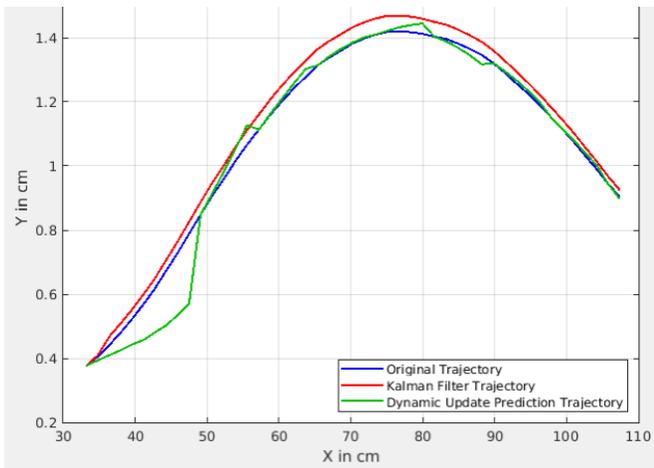

**Figure 4. Dynamic update compared with Kalman Filter for pedestrian #15.**

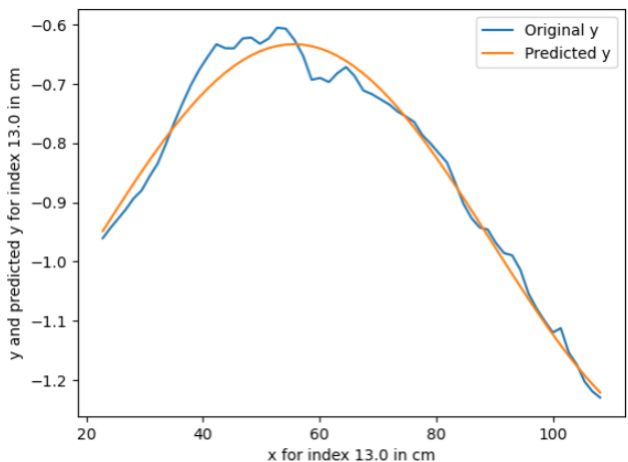

**Figure 5. Symbolic Regression fit for the trajectory of pedestrian #13 in the region of approach. The predicted trajectory is of the form – *0.335\*sin(0.046\*x – 4.137) – 0.968*.**

Given the variability and complexity of human movement, especially in dynamic environments, it is possible that swerving motions may not match the functional forms obtained above because human behavior varies substantially for diverse populations in diverse built environments. In such cases, relying solely on pre-existing data would not suffice. Symbolic Regression is a promising approach to determine the functional form of the swerving trajectory. Unlike traditional regression methods that fit parameters within a predetermined model structure, Symbolic Regression explores a vast space of possible structures to identify the most suitable functional form that describes the observed data.

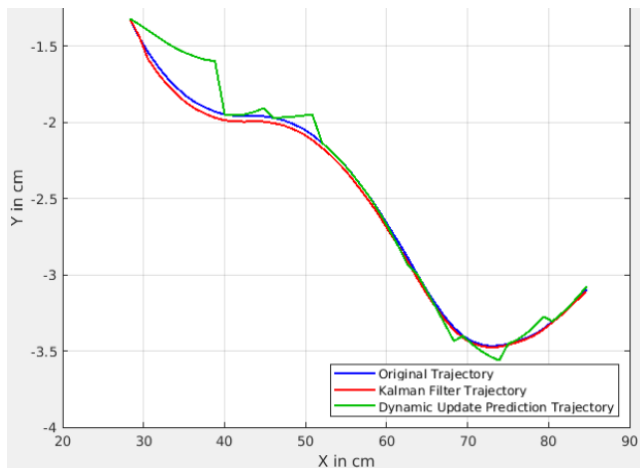

**Figure 6. Dynamic update compared with Kalman Filter for pedestrian #8.**

Before training the symbolic regression model, we provide a set of basic mathematical functions and operators to combine, such as "sin," "exp," "+," and "*". The model then searches for and combines these expressions to find the best-fitting equation for the given data. The symbolic regression tool iteratively refines these combinations, optimizing for both accuracy and simplicity, to find the most suitable mathematical representation of the underlying patterns in the data.

We used the PySR tool to perform symbolic regression to identify functional forms of the region of approach in the collision avoidance maneuvers. We specified "sin," "cos," "*," and "+" as parameters. The tool was able to create a function that fit the collision avoidance trajectory, albeit with overfitting, which made the curves noisier than ideal. To more reasonable functional forms, the curves were approximated by removing insignificant terms. The collision avoidance trajectory in the regions of approach could all then be fitted with a functional form of *k\*sin(a\*x + b) + c*, where *x* is the distance along the direction of motion, and *k, a, b* and *c* are constants. These parameters can then be estimated from the data. Figures 6-8 show the fits with this approach.

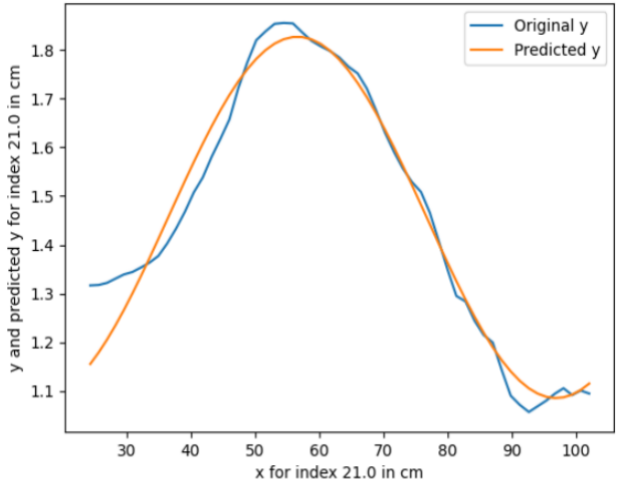

**Figure 7. Symbolic Regression fit for the trajectory of pedestrian #21 in the region of approach. The predicted trajectory is of the form – *0.371\*sin(–0.078\*x – 9.72) + 1.456*.**

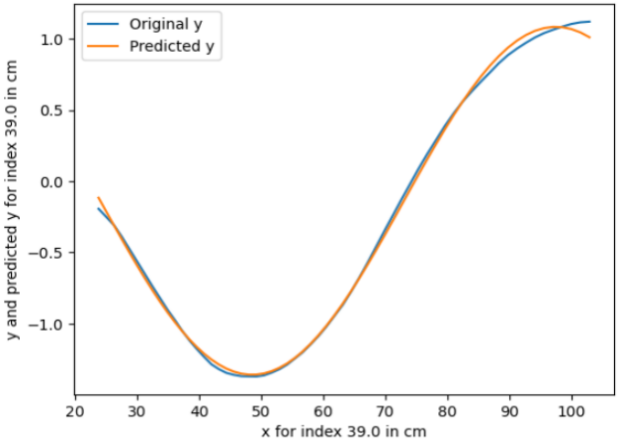

**Figure 8. Symbolic Regression fit for the trajectory of pedestrian #21 in the region of approach. The predicted trajectory is of the form *–0.371\*sin(–0.078\*x – 9.72)+1.456*.**

## 5. DISCUSSION

In order to model swerving, we explored expressing the deviation in terms of different variable choices. The use of *x* and distance *D* to the other colliding pedestrian yielded the simplest forms that worked for all cases studied. While *D* is dependent on y, this is not a problem in developing predictions in practice. In order to predict the trajectory, we break the time interval into a large number of time steps. Using *D(t-Δt)* as an approximation to *D(t)* is sufficiently accurate when Δt is small.

Kalman filter is widely used for data assimilation in physical systems. We wish to explore its effectiveness for human motion. The results show that it is affective for that purpose. However, it lacks (i) the ability to produce different possible scenarios to account for the inherent variation in human behavior and (ii) interpretability to provide insight to the human in the decision-making loop.

The first issue can be addressed through including data assimilation to our swerving model above. We provided a simple data assimilation strategy. It is sufficiently accurate after the initial time interval. Improvements to this would entail better prediction of the first interval and better data assimilation strategies.

The second issue would require identifying new types of behaviors and then providing an interpretation in terms that can be understood by the human in the loop. We performed the first step, using symbolic regression to identify functional forms suitable to express the behavior. Advances in this direction would involve relating such functional forms to physically meaningful human behavior, such as panic.

We next summarize the limitations of this work. We focused exclusively on scenarios where pedestrians approach perpendicularly. Data sets with non-perpendicular approaches are available. Swerving models for those would relate the deviation to the angle of approach too.

Another limitation of this model is that it does not account for simultaneous swerving and speed changes. This is not due to a methodological constraint but rather the nature of the empirical data from the database, which showed minimal variation in speed. This result is consistent with other empirical studies, mentioned earlier, where speed changes are observed only for acute angles of approach. Subsequent models for more general angles of approach can be expected to show variations in speed.

## 6. CONCLUSIONS

We have developed a model for swerving when pedestrians approach each other perpendicularly. This can help adapt social force models to broader contexts such as airports [18]. We have also shown how data assimilation can be used to improve predictions dynamically.

Future work will include generalizing the above work to different approach angles and improving the techniques for data assimilation, including the initial prediction, the actual data assimilation techniques, and interpreting new behaviors. We will also work on the other components required to complete the digital twin, such generating possible scenarios for human response to interventions.

## ACKNOWLEDGEMENTS

This material is based upon work supported by the National Science Foundation under grant nos. 1931511, 2027514, 1931483, and 2027518 and National Institutes of Health under grant no. 1R15LM013382-01A1. Any opinions, findings, and conclusions or recommendations expressed in this material are those of the authors and do not necessarily reflect the views of the National Science Foundation or the National Institutes of Health.

## BIOGRAPHY

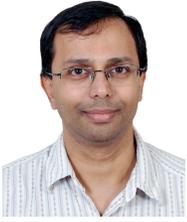

**Ashok Srinivasan** *is the William Nystul Eminent Scholar Chair and Professor of Computer Science at the University of West Florida and a Fulbright Fellow. He obtained a Ph.D. in Computer Science from the University of California, Santa Barbara. His research interests lie in the application of supercomputing to scientific applications and public health policy applications.*

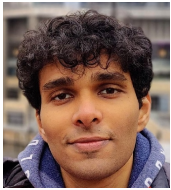

***Satkkeerthi Sriram** received a masters in Computer Science from the University of West Florida and performed this work as a research assistant there.*

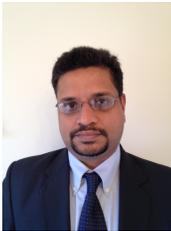

**Sirish Namilae** *is a Professor of Aerospace Engineering at Embry-Riddle Aeronautical University. He is an alumnus of the Indian Institute of Science, where is obtained an M.E. in Materials Science, and he has a Ph.D. from Florida State University. His research interests are in the areas of particle dynamics, multiscale modeling, and composite materials.*

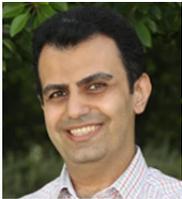

***Andrew Mahyari** is a Research Scientist at the Institute for Human Machine Cognition and Adjunct Professor at the University of West Florida. He has a PhD in Electrical, Electronics, and Communications Engineering from Michigan State University. His research interests lie in machine learning applications, including computer vision, reinforcement learning, deep learning, and deep generative.*